\documentclass[conference]{IEEEtran}
\IEEEoverridecommandlockouts

\usepackage{url}

\usepackage{cite}
\usepackage{amsmath,amssymb,amsfonts}
\usepackage{algorithmic}
\usepackage{graphicx}
\usepackage{textcomp}
\usepackage{xcolor}
\usepackage{subfigure}
\usepackage{array}
\usepackage[T1]{fontenc}
\usepackage[latin9]{inputenc}
\newcolumntype{P}[1]{>{\centering\arraybackslash}p{#1}}
\newcolumntype{M}[1]{>{\centering\arraybackslash}m{#1}}
\def\BibTeX{{\rm B\kern-.05em{\sc i\kern-.025em b}\kern-.08em
    T\kern-.1667em\lower.7ex\hbox{E}\kern-.125emX}}
\begin{document}

\title{DiffLoop: Tuning PID Controllers by Differentiating Through the Feedback Loop}

\author{\IEEEauthorblockN{Athindran Ramesh Kumar}
\IEEEauthorblockA{\textit{Department of Electrical Engineering} \\
\textit{Princeton University}\\
Princeton, USA\\
arkumar@princeton.edu}
\and
\IEEEauthorblockN{Peter J. Ramadge}
\IEEEauthorblockA{\textit{Department of Electrical Engineering} \\
\textit{Princeton University}\\
Princeton, USA\\
ramadge@princeton.edu}
}

\maketitle

\begin{abstract}
Since most industrial control applications use PID controllers, PID tuning and anti-windup measures are significant problems. This paper investigates tuning the feedback gains of a PID controller via back-calculation and automatic differentiation tools. In particular, we episodically use a cost function to generate gradients and perform gradient descent to improve controller performance. We provide a theoretical framework for analyzing this non-convex optimization and establish a relationship between back-calculation and disturbance feedback policies. We include numerical experiments on linear systems with actuator saturation to show the efficacy of this approach.
\end{abstract}

\section{Introduction}
PID controllers are the most popular form of feedback control in industrial applications \cite{aastrom2010feedback}. 
In general, the PID gains need to be tuned to obtain good performance. In addition,
potential actuator saturation must be taken into account, since saturation can induce 
integrator wind-up, resulting in unexpected transients during operation. 
To address the above issues,
PID tuning \cite{cominos2002pid} can be performed using classical control based on a model of the plant devised either from prior knowledge or system identification. 
 Model-free \cite{ziegler1942optimum, skogestad2001probably} PID tuning has also been explored by performing selective experiments to tune the gains gradually. 
Notably, machine learning approaches such as \cite{zhou1992pid,lawrence2020optimal,mitsukura1997genetic} tune the controller parameters without a system model.
In practice, we usually have some prior knowledge of the system and can use it to obtain a coarse system model. The results in \cite{recht2019tour} show that a coarse model can be useful in obtaining an initial robust controller.
Indeed, given a coarse model for a linear system, 
we can apply several controller design techniques, both classical (PID, loop shaping) and modern (LQR, $H^{\infty}$).
However, when non-linearities induced by actuator saturation are present, simple controller design techniques are not available.
In this setting, the back-calculation method \cite{torstensson2013comparison,aastrom2010feedback} is the simplest anti-windup technique. 
\par 
We explore using the back-calculation method 
together with a non-convex optimization approach based on differentiating through the system model, actuator, and the feedback loop to tune the feedback gains. This approach is inspired by recent work on differentiable physics engines/models \cite{degrave2019differentiable,de2018end}. 
In particular, we episodically tune the controller parameters by simulating with the current parameters, evaluating the cost, and performing gradient descent on the cost objective. 
By propagating the gradients over time for the entire simulation, we capture the controller parameters' long-term dependencies on the system's dynamics. 
The use of automatic differentiation tools, e.g., in TensorFlow \cite{tensorflow2015-whitepaper}, PyTorch \cite{paszke2017automatic}, or as stand-alone code \cite{Autograd, jax2018github}, enable us to easily compute required gradients on-the-fly without using analytical techniques. 
The computational cost scales with the simulation horizon. For reasonable time horizons, the method can be efficiently implemented on a modern CPU.
\par
The non-convex optimization problem of interest
is posed as an output-feedback controller design with augmented state and marginally stabilizable, detectable dynamics. 
This framework is developed by treating the errors due to actuator saturation as an additional disturbance and predicting future disturbances using disturbance estimates at previous time-steps. 
Using this approach, we can show an equivalence between the back-calculation method and the class of disturbance feedback controllers introduced in \cite{agarwal2019online}. 
We leave the convergence analysis of gradient descent for this optimization problem to future work. 
Here we perform simulations on four different systems with saturation to show the efficacy of the approach.
\par
We discuss our work's relation to the existing literature in section \ref{sec:rwork}.  Then in section \ref{sec:disturbance}, we introduce a framework to analyze the optimization's convergence properties.
We perform numerical experiments to illustrate our approach in section \ref{sec:numerical}, and conclude in section \ref{sec:conclusion}.
\section{Related Work}\label{sec:rwork}
\subsection*{Machine learning for PID tuning and anti-windup design}
PID tuning is a well-studied problem, see e.g., \cite{aastrom1995pid, ziegler1942optimum, skogestad2001probably}.  
Much of the previous work has focused on black-box optimization, reinforcement learning, and model-free tuning. We briefly discuss these approaches below.
\par
Black-box PID tuning includes approaches such as genetic algorithms \cite{mitsukura1997genetic,herrero2002optimal}, particle swarm optimization \cite{chen2007particle}, and model-free  decision trees \cite{zhou1992pid}. These do not incorporate an explicit model of the system. A distinct approach to PID tuning uses reinforcement learning. 
Here, both model-based and model-free approaches are explored. For example,
the authors of \cite{doerr2017model} use a model-based reinforcement learning approach with a Gaussian Process model to tune the PID gains for a seven degree-of-freedom robot arm. However, their algorithm ignores actuator saturation.  Reference \cite{boubertakh2010tuning} tunes fuzzy PD and PI controllers using the Q-learning algorithm.  
 Similarly, \cite{shi2018adaptive} uses the model-free Q-learning algorithm to tune the PID gains for a cart-pole system.  
Also, \cite{lawrence2020optimal} uses an actor-critic RL algorithm with a neural Q-function to tune the PID parameters with back-calculation.
%
 \par
In addition to tuning the PID parameters, we want to address actuator saturation. Anti-windup compensation has been tackled by both classical \cite{bohn1995analysis, torstensson2013comparison,aastrom2010feedback} and modern control \cite{galeani2009tutorial} techniques. 
Among these approaches, the back-calculation method \cite{aastrom2010feedback} is a simple and effective scheme for compensating for integrator wind-up. It requires 
an actuator model with knowledge of the saturation limits. Even though the scheme lacks the formal stability and robustness guarantees of modern anti-windup design, its simplicity is appealing.  
Also relevant is the idea of 
treating the errors due to actuator saturation as a disturbance \cite{kheirkhahan2017robust}. We adopt this approach.
\subsection*{Differentiable models}
There is recent work in building physics engines that can be differentiated through to update model parameters or train controllers \cite{chang2016compositional,degrave2019differentiable,de2018end,amos2017optnet}. 
Notably, in \cite{de2018end}, the differentiable model approach is effective even in learning controllers from raw images. 
Our work is complementary 
in two aspects. First, before extending the differentiable physics approach to complex problems, it would be interesting to understand its performance on simple systems with purely saturation 
non-linearities.
If this is effective, then second, 
it would be of interest to tune PID controllers (with saturation) using standard automatic differentiation tools.
\subsection*{Theoretical machine learning and control}
Recent growth at the intersection of machine learning theory and control theory has enabled better understanding of some long-standing problems. 
A class of disturbance feedback policies is introduced and optimized using a convex relaxation in \cite{agarwal2019online, hazan2020nonstochastic,simchowitz2020improper}. 
In \cite{recht2019tour}, it has been shown that model-based control approaches are often sample-efficient compared to model-free reinforcement learning for linear systems control. 
References \cite{fazel2018global,zhang2020policy}, show the convergence of gradient descent and policy optimization for non-convex state feedback LQR and $H_{2},$ $H_{\infty}$ controller design. 
However, the output feedback case is less studied. 
In our work, we seek a model-based approach applicable to systems with saturation. Actuator saturation is one of the most common forms of non-linearity in practical control systems. We explore a  relationship between the back-calculation anti-windup method and disturbance feedback policies.
\section{Disturbance feedback for anti-windup compensation}\label{sec:disturbance}
Assume the system to be controlled has a stabilizable and detectable state space representation:
\begin{equation}
\begin{aligned}
x_{t+1}&=Ax_{t}+Bu_{t}+w_{t}\label{eq:linear1}\\
y_{t}&=Cx_{t}+e_{t} .
\end{aligned}
\end{equation}

Here $u_t\in \mathbb{R}^{m}$, $x_t\in\mathbb{R}^{n},$ and $y_t\in\mathbb{R}^{p}$ are the input, state, and output at time $t,$ respectively.
To model  actuator saturation, we modify \eqref{eq:linear1} to:
\begin{align}
x_{t+1}&=Ax_{t}+B\text{sat}(u_{t})+w_{t} \label{eq:linear3}
\end{align}
\subsection{Back-calculation method}
In the back-calculation method \cite{aastrom2010feedback}, the errors due to actuator saturation are integrated and fed back to prevent windup.
Let $r_{t}$ denote the reference signal to be tracked, and
$P_t,$ $I_t, $ and $D_t$ denote the proportional, integral, and derivative signal components of the PID controller, respectively.  Then a standard back-calculation PID controller  is given by:
\begin{equation}
\begin{aligned}
 P_t& = k_p\left(r_{t}-y_{t}\right)\\
\!\!\!  \!D_t&= \alpha D_{t-1} + k_{d}\Delta y_t\\
I_{t+1}&=I_t \!+\!k_{i}\left(r_{t}\!-\!y_{t}\right)\!+\!b(\text{sat}(v_t)\!-\!v_t)\\
v_t&=P_t+I_t+D_t\\
\text{sat}(v_t)&=\text{clamp}(v_t, u_{\textrm{low}},u_{\textrm{high}}).
\end{aligned}
\end{equation}

Here $\Delta$ is the difference operator, 
$\alpha $ is a filter parameter, 
$k_{p}$, $k_{i}$, $k_{d}$ and $b$ are the proportional, integral, derivative and back-calculation feedback gains. Finally,  $u_{\textrm{low}}$ (resp. $u_{\textrm{high}}$) is the minimum (resp. maximum) actuator output. Here, $r_{t}$ is just a reference operating point and can be set to $0$ for analysis. 
\par
\subsection{Disturbance feedback policies and back-calculation}
We now connect the back-calculation technique to disturbance feedback policies. To do so, we formulate a  controller design starting from the linear state-space system in \eqref{eq:linear1}.

\subsubsection*{PID controller design}
 We first pose PID tuning as output-feedback controller design. Since we feedback integral and derivative terms, we append these terms to the state. Let
 \begin{equation}
\begin{aligned}
i_{t+1}&=\sum_{t^{'}=1}^{t+1} x_{t^{'}}=i_{t}+x_{t}\\
d_{t+1}&=x_{t}-x_{t-1}.
\end{aligned}
 \end{equation}
Then, form the augmented state 
$X_{t}=[x_{t};x_{t-1};i_{t}],$ and 
the augmented state-space equations:
\begin{equation}
\begin{aligned}
\begin{bmatrix}x_{t+1}\\x_{t}\\i_{t+1}\end{bmatrix}
&= \begin{bmatrix}
A&\mathbf{0}&\mathbf{0}\\
I&\mathbf{0}&\mathbf{0}\\
I&\mathbf{0}&I
\end{bmatrix}
\begin{bmatrix}x_{t}\\x_{t-1}\\i_{t}\end{bmatrix}
+
\begin{bmatrix}B \\ \mathbf{0}\\ \mathbf{0}\end{bmatrix} u_{t} +w_{t}\\
Y_{t}& = \begin{bmatrix}
C&\mathbf{0}&\mathbf{0}\\
\mathbf{0}&\mathbf{0}&C\\
C&-C&\mathbf{0}
\end{bmatrix}
\begin{bmatrix} x_{t}\\x_{t-1}\\i_{t} \end{bmatrix}+e_{t}.
\end{aligned}
\end{equation}
We write these equations more concisely as:
\begin{equation}
\begin{aligned}
X_{t+1}& = \tilde{A}X_{t} + \tilde{B}u_{t} + w_{t}\label{eq:asss1}\\
Y_{t}& = \tilde{C}X_{t}+e_{t},
\end{aligned}
\end{equation}

with $w_{t}$ and $e_{t}$ defined appropriately. 
If the initial system is stabilizable and detectable, so is the augmented version. This is because the dynamics of the augmented states are not directly controllable. If the system in equation \ref{eq:linear1} is exponentially stable, then the augmented states cannot grow exponentially.

All PID controllers (with $\alpha=0$) can be expressed as $u_{t}=-KY_{t}$ for the system. If the state was measurable, i.e. $\tilde{C}=I$, the problem would reduce to LQR, and the optimal PID gains can be obtained by both gradient descent and policy optimization \cite{fazel2018global}. 
However, the output-feedback controller optimization problem is still open
\eqref{eq:asss1}. For $\alpha\neq 0$, it is straightforward to introduce a controller that stores the filtered derivative as its state. Such a controller would not be purely state/output feedback. Let $X^{c}_{t}$ be the controller state at time $t$. The controller with $\alpha\neq 0$ can be expressed as:
\begin{equation}
    \begin{aligned}
X^{c}_{t+1}&=\alpha X^{c}_{t}+K_{b}Y_{t}\\
u_{t}&=-K_{x}X^{c}_{t}-KY_{t}
    \end{aligned}
\end{equation}
We append $X_{t}^{c}$ to $X_{t}$ to augment the state-space further:
\begin{equation}
    \begin{aligned}
    \begin{bmatrix}
     X_{t+1}\\X^{c}_{t+1} 
    \end{bmatrix}
    &= \begin{bmatrix}
     \tilde{A}& \mathbf{0}\\ \mathbf{0}& \alpha I 
    \end{bmatrix}\begin{bmatrix}
     X_{t}\\X^{c}_{t}\end{bmatrix}+\begin{bmatrix}
     \tilde{B} & \mathbf{0}\\ \mathbf{0} & K_{b}
     \end{bmatrix} \begin{bmatrix} u_{t}\\ Y_{t} \end{bmatrix} + \begin{bmatrix} w_{t} \\ \mathbf{0} \end{bmatrix}\\
     Y_{t}& = \tilde{C}X_{t}+e_{t}\\
    u_{t}&=-K_{x}X^{c}_{t}-KY_{t}
    \end{aligned}
\end{equation}
Henceforth, we assume $\alpha = 0$ with the extension to $\alpha \neq 0$ possible with the above controller state.
\subsubsection{Actuator saturation as a disturbance}
Let $w^a_t \triangleq B' (\text{sat}(u_t) - u_t)$ denote the saturation error.  
We treat the saturation error as a disturbance and write
\begin{equation}
    \begin{aligned}
   X_{t+1}&=\tilde{A}X_{t}+\tilde{B}\text{sat}(u_{t})+w_{t}\\
        &=\tilde{A}X_{t}+\tilde{B}u_{t}+w^{a}_{t}+w_{t} .
    \end{aligned}
\end{equation}
The error $w^{a}_{t}$ is a non-linear function of the input.  
It can be modeled as adversarial (as in some online learning settings). To handle adversarial disturbances, \cite{agarwal2019online} introduces disturbance feedback policies of the form:
\begin{align}
       u&=-KX_{t}-\sum_{l=1}^{h}K^{[l]}_{d}w_{t-l}.
\end{align}
If this approach is provided with a stabilizing controller $K$, the resulting online optimization is convex and provides tight regret bounds\cite{agarwal2019online}.
To draw the connection between disturbance feedback policies and back-calculation, we observe that if $h$ is the length of the simulation horizon and $K_{d}^{[l]}=K_{d}$ for all $l$, this class of controllers reduces to the back-calculation method.
In back-calculation we integrate the disturbances due to actuator saturation and feed it back to the input.

\subsubsection{Disturbance feedback policies in episodic learning}
Here, we focus on an episodic setting. We run an episode with the PID parameters and tune the parameters episodically. In this case, additional modeling assumptions are required for disturbance feedback controllers to be meaningful. 
We postulate that the disturbance has marginally stable dynamics that we want to learn. Hence, we introduce a predictor for $w_{t}^{a}$:
\begin{align}
w_{t}^{a}=\sum_{i=1}^{h}M^{[i]}w_{t-i}^{a} .
\end{align} 
We want the state to be sufficient for selecting  the control action at any given time. Hence we augment the state to 
$$Z_{t}=[\tilde{X}_{t};w_{t}^{a};w_{t-1}^{a};w_{t-2}^{a}\dots;w_{t-h}^{a}].$$
When the disturbance is purely stochastic (no internal dynamics), state/output feedback
is optimal. 
Here we model disturbance dynamics and use the model to obtain a class of disturbance feedback policies. 
Using the augmented state $Z_{t}$, we can write the dynamics as:

\begin{align*}
Z_{t+1} & =  \begin{bmatrix}
\tilde{A} & I & 0 & 0\\
0 & M^{[1]} & M^{[2:h-1]} & M^{[h]}\\
0 & I & I & 0
\end{bmatrix} Z_{t} 
+ \begin{bmatrix}
\tilde{B}\\
0\\
0
\end{bmatrix} u_{t} + w^{r}_{t}\\
Y^{z}_{t} &  =
\begin{bmatrix}
\tilde{C} & \mathbf{0}\\
\mathbf{0} & I
\end{bmatrix} Z_{t} + e^{r}_{t}
\end{align*}
with $w^{r}_{t}$ the unmodeled disturbance. The dynamics of $w_{t}^{a}$ has to be at least marginally stable in order for a controller to stabilize the system, but it need not be asymptotically stable.  
The class of output-feedback controllers $u_{t}=-KY^{z}_{t}$ can be expressed as:
\begin{equation}
    \begin{aligned}
u_{t} & =-K_{c}Y_{t}-K_{d}^{'}w_{t:t-h}^{a}\\
 & =-K_{c}Y_{t}-K_{d}^{'}\left[\begin{array}{cc}
M^{[1:h]} ;& I\end{array}\right]w_{t-1:t-h}^{a} \nonumber \\
 & =-K_{c}Y_{t}-K_{d}w_{t-1:t-h}^{a} .
    \end{aligned}
\end{equation}
This gives 
a general class of controllers that includes disturbance-feedback controllers and the back-calculation method. Further, for the specific scenario of anti-windup compensation, we can recover 
$w_{t}^{a}$ required in the policy using the assumed actuator model.
\subsubsection{Optimization for Parameter Tuning}
In order to tune $K_{c}$ and $K_{d}$, we 
perform gradient descent with the objective function
\begin{align}
\min_{K_{c},K_{d}} & \sum_{t} y_{t}^{T}Qy_{t}+u_{t}^{T}Ru_{t}.
\end{align}
Solving this optimization problem using gradient-descent does not require explicit learning of the disturbance dynamics 
since the controller parameters can be directly optimized. Note that the dynamics
are only marginally stabilizable. However, the uncontrollable and marginally stable components of the state do not appear in the cost function. A theoretical understanding of gradient descent for marginally stabilizable systems with output feedback is beyond the scope of this work. We relegate this study to future work and for the present paper perform an empirical evaluation of this non-convex optimization problem.

\section{Numerical Results}\label{sec:numerical}
We illustrate the proposed approach of tuning the controller parameters by differentiating through the model around the feedback loop
using four different systems.
The setup for the different experiments is 
summarized in Table \ref{tab:exp}.
\begin{table*}[h]
\begin{center}
\begin{tabular}{|M{0.1\textwidth}|M{0.28\textwidth}|M{0.1\textwidth}|M{0.1\textwidth}|M{0.25\textwidth}|}
\hline
& Plant & Actuator limits & Limits on the step reference & Initial feedback gains\\
\hline
System 1&$P(s)=\frac{2e^{-0.02s}}{s-0.995}$&$\pm 3.3$&$\pm 4$&$k_{p}=4$, $k_{i}=10$, $b=0.5$\\
\hline
System 2&$P(s)=\frac{1}{(20s^{2}+10s+1)}$&$\pm 7.0$&$\pm 4$&$k_{p}=10$, $k_{i}=1.5$,$k_{d}=8$, $b=0.4$\\
\hline
System 3&$P(s)=\frac{1}{(s+0.1)(s-0.1)}$&$\pm 3.0$&$\pm 2.9$&$k_{p}=20$, $k_{i}=2$, $k_{d}=5$, $b=1$\\
\hline
System 4&$P(s)=\frac{(s+0.5)(s+0.3)}{(s+0.1)(s+0.2)(s+0.4)(s+0.6)}$&$\pm 4.0$&$\pm 3$&$k_{p}=20$, $k_{i}=8$, $k_{d}=10$, $b=0.2$\\
\hline
\end{tabular}
\end{center}
\caption{Setup of the experiments}\label{tab:exp}
\end{table*}
\par
In each of the experiments, we use a linear system with actuator saturation. Systems 1,3 are unstable, and systems 2,4 are stable. The systems are progressively higher order and complex.  The only non-linearity in the system is actuator saturation. The saturation limits were chosen to induce windup in the absence of any anti-windup strategy. %
 The simulation horizon is chosen to balance the steady-state cost with the transient costs to obtain a good step response. If the simulation horizon is too short, then the costs in the loss function due to the transients will dominate. The controllers have to ensure precise convergence to the step after the transients.
We convert the continuous-time linear system into its discrete-time counterpart using zero-order hold (ZOH) and simulate the system in discrete time. We tune the feedback gains episodically by performing gradient descent on the squared-error cost function $\sum_{t=1}^{T}(y_{t}-r_{t})^{2}.$ 
We generate $30$ reference signals and segregate them into $20$ signals used for training and $10$ used for testing. 
\par
For each system, we compare four distinct controllers:  (1) The initial PI/PID controller is tuned using classical control techniques to work well without actuator saturation.  Integrator windup impacts performance when saturation is present; 
(2) We initialize a back-calculation constant manually to decrease windup;
(3) The PID and back-calculation parameters are then optimized using gradient descent with the Adam optimizer \cite{kingma2014adam} to obtain a third controller; 
(4) This controller has dynamically changing feedback gains modeled as a small neural network with tracking error and actuation error as inputs.
\par
\begin{figure}
   \centering
    \includegraphics[scale=0.43]{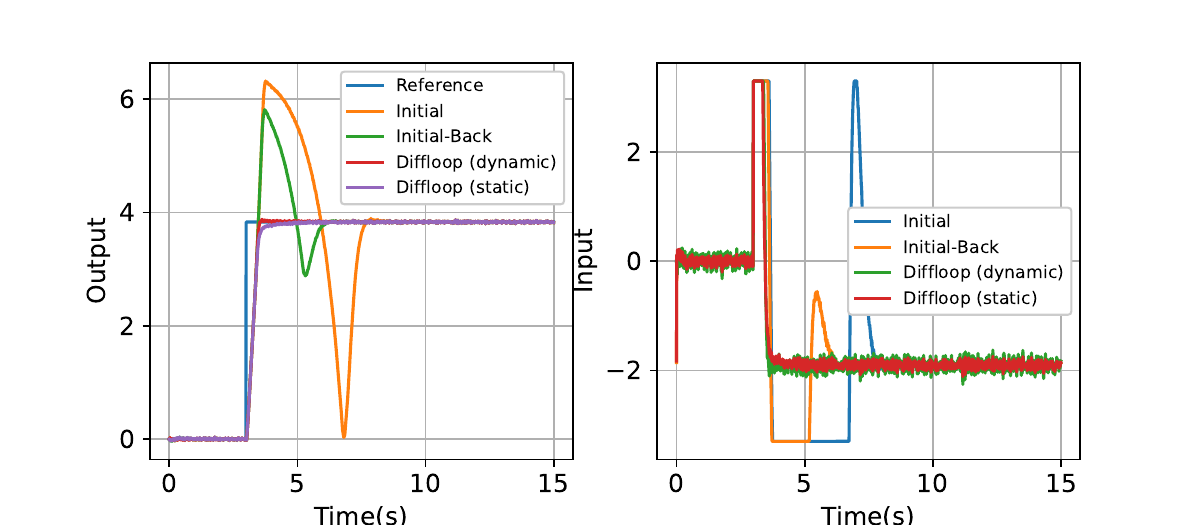}
    \caption{Performance of the four controllers on a difficult test reference for system 1. Left: Output of the SISO system with the four controllers. Right: Input to the system with the four controllers. See section \ref{sec:numerical} for interpretation.}
    \label{fig:founstable1}
\end{figure}
\begin{figure}
    \includegraphics[scale=0.2]{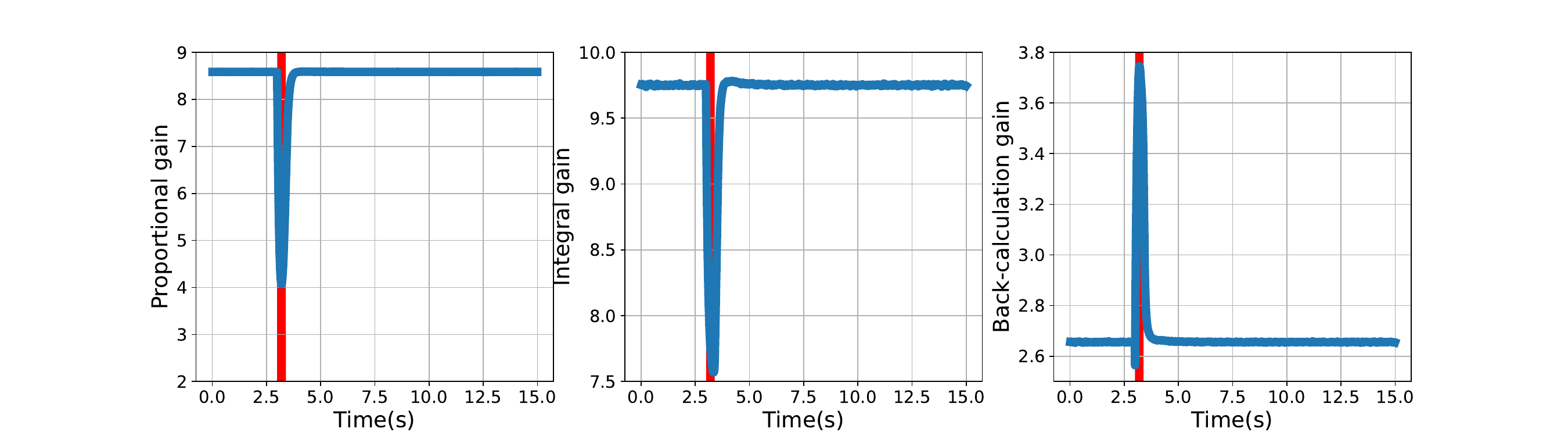}
    \caption{Variation of the tuning constants with time for the dynamic PI controller for system 1. The red shaded region indicates periods of saturation. The controller gains switch during saturation.}
    \label{fig:founstable2}
\end{figure}

Simulation 1 performs step-reference tracking on system 1.  After optimization, the parameters converged to $k_{p}^{*}=16.58$, $k_{i}^{*}=11.07$ and $b^{*}=0.87$. 
The cost on the training and testing reference signals of the four different controllers are summarized in Table \ref{tab:sys1}.
Optimization of the feedback gains using gradient descent is effective in improving performance. 
Using a neural-network to change the feedback gains dynamically does not further improve performance. This suggests that  simple controllers are sufficient to control linear systems with saturation.
From Figure \ref{fig:founstable1}, we see that both the dynamic and static optimized PI controllers are effective in preventing windup and tracking the reference accurately.
In Figure \ref{fig:founstable2}, we plot the variation of the dynamic PI controller's feedback gains with time. 
It is interesting that the feedback gains switch during the transition and return to the initial values. 
This indicates that switching between multiple controllers depending on the input is an effective strategy to improve performance. 
However, for the simple systems used here, 
the static controller is sufficient for good performance.
\begin{table}
\begin{center}
\begin{tabular}{|M{0.2\textwidth}|M{0.1\textwidth}|M{0.1\textwidth}|}
 \hline
 Method & Training cost & Test cost \\ 
 \hline
 Initial PI &$304.4\pm432.3$&$349.0\pm503.3$\\  
 \hline
 Initial PI with backcalculation &$178.2\pm153.0$&$189.0\pm164.8$\\ 
 \hline
 PI+backcalculation optimized &$110.2\pm79.8$&$114.7\pm82.3$\\
 \hline
 Dynamic PI+backcalculation optimized&$109.5\pm79.9$&$114.0\pm82.2$\\
 \hline
\end{tabular}
\end{center}
\caption{Squared error cost of the 
four controllers on system 1. See section \ref{sec:numerical} for interpretation.}\label{tab:sys1}
\end{table}
\par
\begin{figure*}
\centering
\subfigure[System 2]{\includegraphics[width=5.1cm]{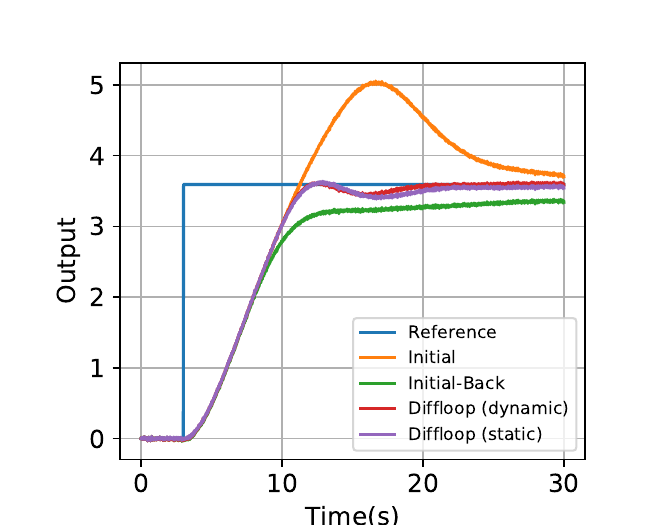}}
\subfigure[System 2 feedback gains]{\includegraphics[width=7.5cm]{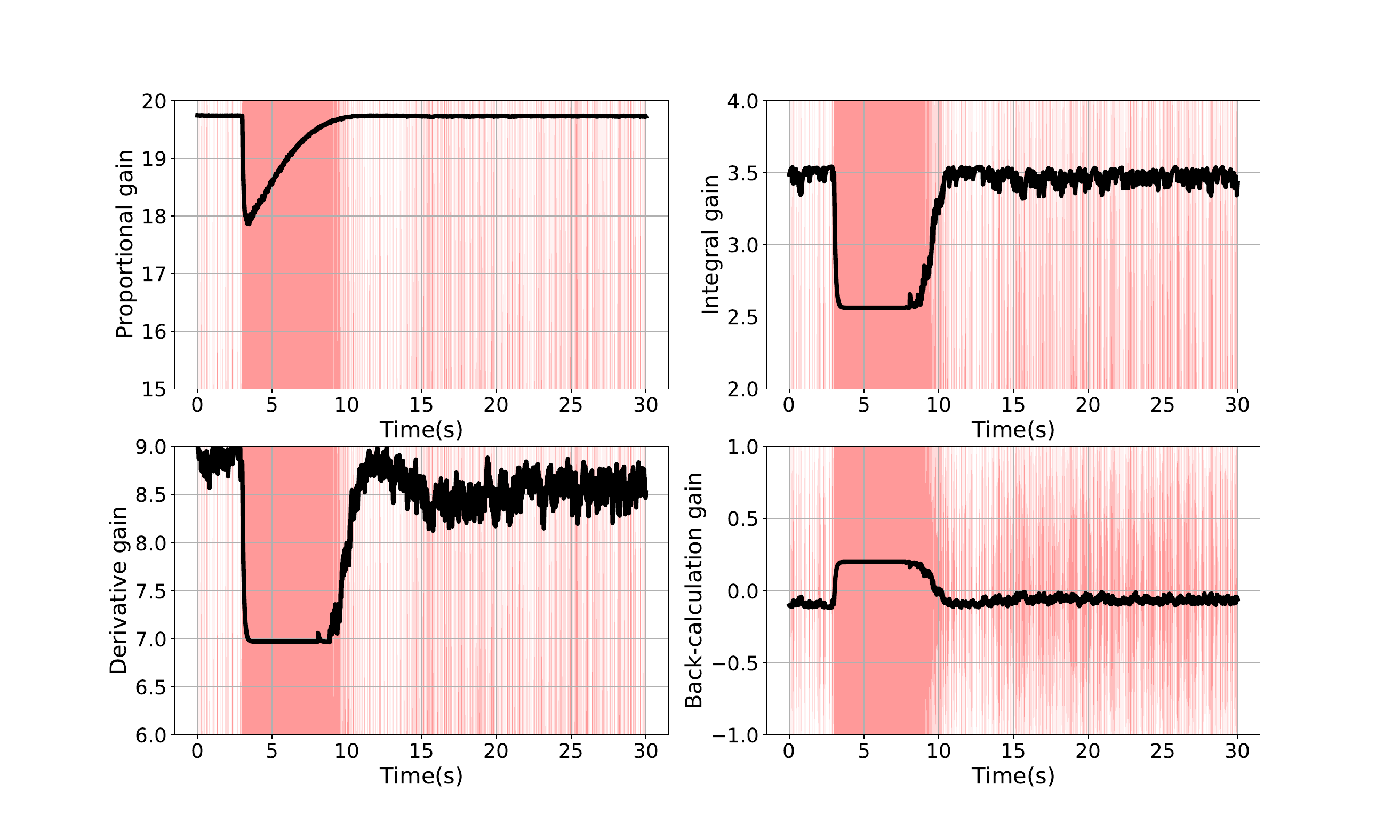}}
\caption{(a) Output of the four controllers on a step input for system 2. We can see that optimizing the PID gains both in the static and dynamic case is effective in providing a good step response. (b) Variation of the feedback gains with time for the Dynamic PID controller. The gains switch during the transition and return to their initial values. The red shaded regions denote periods of saturation}\label{fig:sys33}
\subfigure[System 3]{\includegraphics[width=5.1cm]{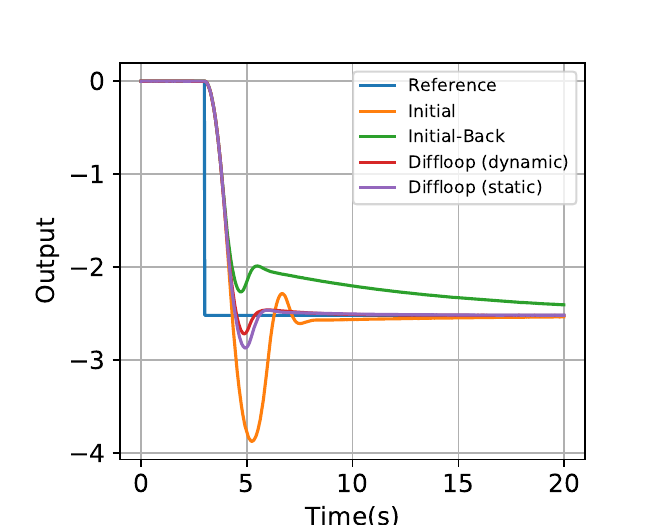}}
\subfigure[System 3 feedback gains]{\includegraphics[width=7.5cm]{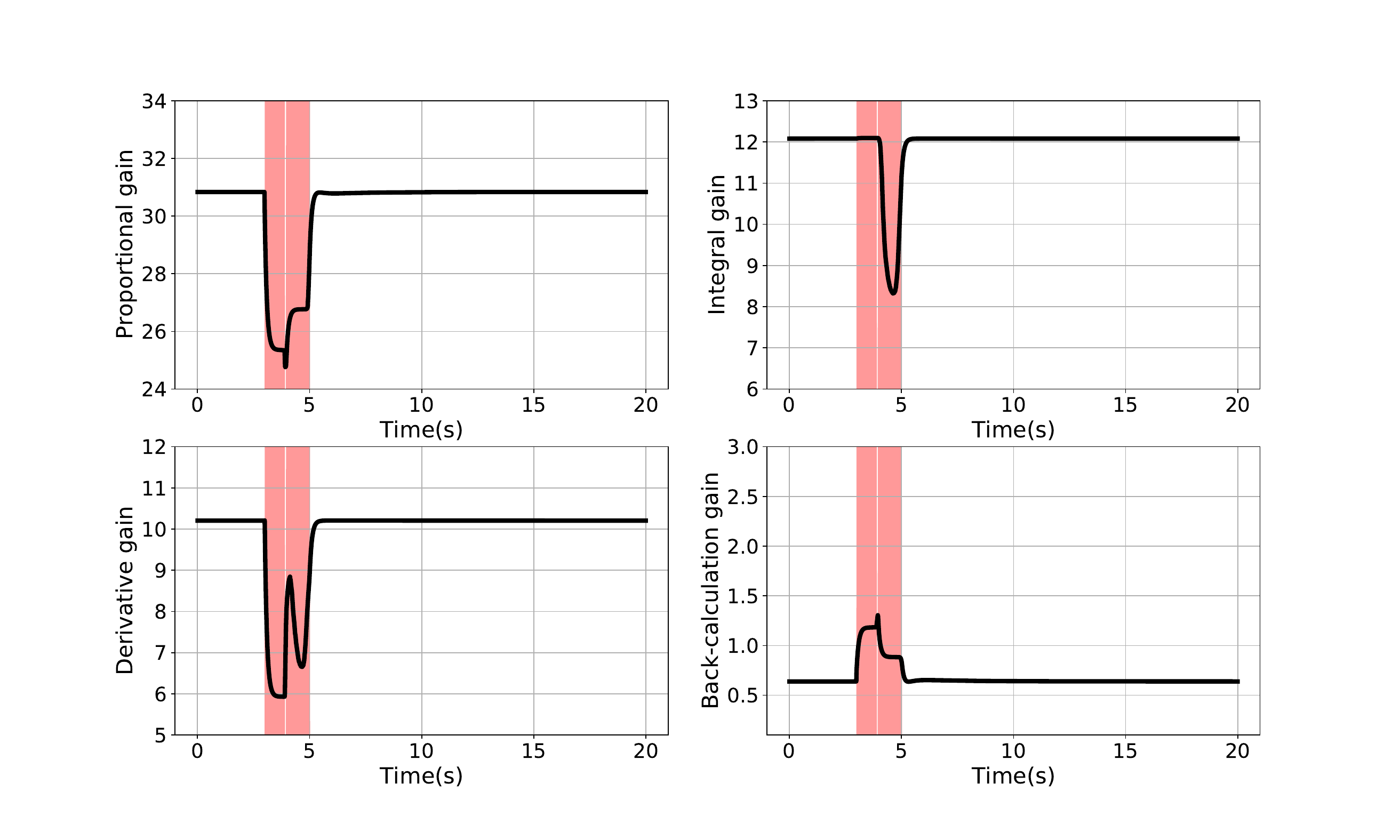}}
\caption{(a) Output of the four controllers on a step input for system 3. We can see that optimizing the PID gains both in the static and dynamic case is effective in providing a good step response. (b) Variation of the feedback gains with time for the Dynamic PID controller. The gains switch during the transition and return to their initial values. The red shaded regions denote periods of saturation}\label{fig:sys34}
\subfigure[System 4 output]{\includegraphics[width=5.1cm]{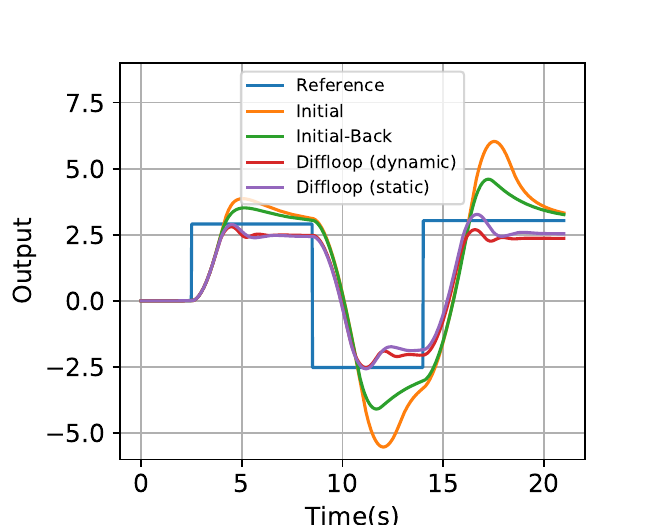}}
\subfigure[System 4 feedback gains]{\includegraphics[width=7.5cm]{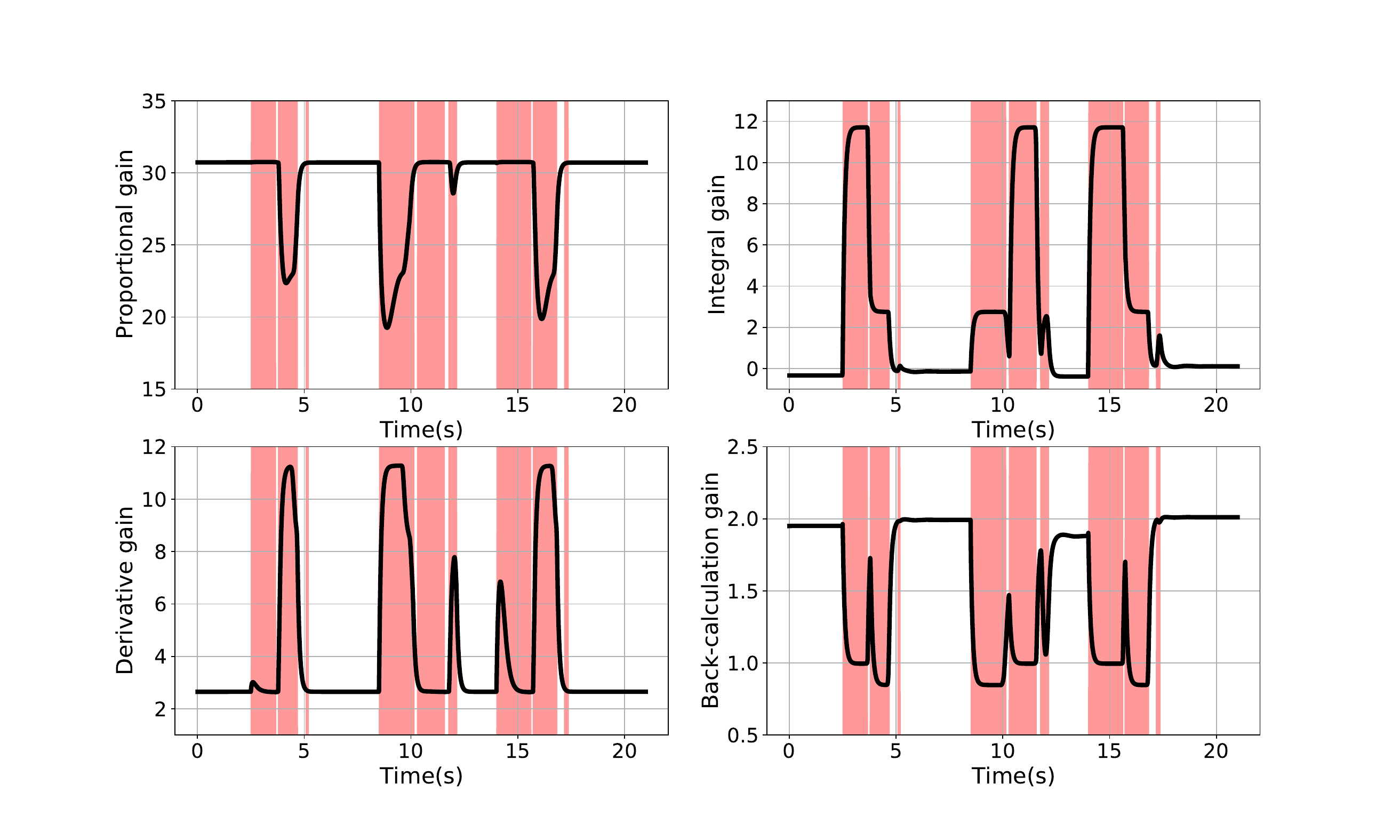}}
\caption{(a) Output of the four controllers on a rapidly switching reference for system 4. We can see that optimizing the PID gains both in the static and dynamic case is effective to some extent to cope up with the hard saturation limit. (b) Variation of the feedback gains with time for the Dynamic PID controller. The gains switch during the transitions and return to their initial values. The red shaded regions denote periods of saturation}\label{fig:sys4}

\end{figure*}

Simulations 2 and 3 repeat the above procedure  on the second-order systems 2 and 3. 
In simulation 2, the parameters converged to $k_{p}^{*}=11.31$, $k_{i}^{*}=1.71$, $k_{d}^{*}=4.26$ and $b^{*}=0.23$. In simulation 3, the parameters converged to $k_{p}^{*}=7.72$, $k_{i}^{*}=1.47$, $k_{d}^{*}=3.01$ and $b^{*}=0.66$. The performance of the controllers in terms of squared error cost is summarized in table \ref{tab:sys23}. Figures \ref{fig:sys33} and \ref{fig:sys34}, indicate that optimization of the PID controllers is effective.
\begin{table}
\begin{center}
\begin{tabular}{|M{0.2\textwidth}|M{0.1\textwidth}|M{0.1\textwidth}|}
 \hline
 Method & System 2 Test cost & System 3 Test cost\\ 
 \hline
 Initial PID & $2115.2\pm1822.9$ &$264.8\pm220.0$\\  
 \hline
 Initial PID with backcalculation & $1788.1\pm1506.7$ &$247.6\pm197.0$\\
 \hline
 PID+backcalculation optimized & $1688.9\pm1405.9$ &$198.2\pm148.7$\\
 \hline
 Dynamic PID+backcalculation optimized& $1686.7\pm1410.7$ &$198.1\pm150.0$\\
 \hline
\end{tabular}
\end{center}
\caption{Performance of the four controllers in terms of squared error for system 2 and system 3. See section \ref{sec:numerical} for interpretation.}\label{tab:sys23}
\end{table}

Finally, in simulation 4, we track a rapidly switching reference using a PID controller with back-calculation. The saturation limits
constrain the controller from switching the outputs rapidly to achieve accurate tracking.
However, the optimization minimizes the squared error cost. Interestingly, for rapidly switching reference signals, the feedback gains converges to $k_{p}^{*}=11.78$, $k_{i}^{*}=-0.47$, $k_{d}^{*}=3.46$ and $b^{*}=0.22$. The integrator gain consistently converges to negative values. 
For a rapidly switching reference, the transient costs outweigh the cost due to lack of precise convergence and noise. Hence, it is better to choose feedback gains that reduce overshoot and windup. This observation gives rise to an interesting phenomenon.  In figure \ref{fig:sys4}, we plot the system's outputs with the four different controllers. Even though actuator saturation 
is too limiting to switch rapidly, the optimized controllers achieve lower squared error costs and perform better. 
\section{Conclusions and Future Work}\label{sec:conclusion}
We outline a PID tuning approach for linear systems with input saturation. This approach differentiates through the model and around the feedback loop to tune the controller parameters. The numerical experiments demonstrate the efficacy of this approach.
We also propose a theoretical framework to analyze the convergence properties for this optimization. However, a convergence proof is beyond the scope of this work. We noted that the framework shows the equivalence of the back-calculation method and disturbance feedback policies.
\par
Future work can extend this technique for generating robust controllers for MIMO systems using robust optimization. Further, the automatic differentiation technique could also be used for tuning PID controllers in robotic systems. 
Output feedback controller optimization also warrants further theoretical study.
\bibliographystyle{plain}
\bibliography{refs}
\end{document}